\documentclass[12pt,fleqn]{article}
\usepackage{epsfig}

\newcommand{\be}{\begin{eqnarray}}
\newcommand{\ee}{\end{eqnarray}}
\newcommand{\nee}{\nonumber\end{eqnarray}}
\newcommand{\nn}{\nonumber}
\newcommand{\noi}{\noindent}
\newcommand\sfrac[2]{{\textstyle \frac{#1}{#2}}}

\newcommand{\eq}[1]  {\mbox{(\ref{eq:#1})}}
\newcommand{\fig}[1] {\mbox{Fig.~\ref{fig:#1}}}

\newcommand{\sect}[1] {\mbox{Section~\ref{sect:#1}}}

\newcommand{\app}[1] {\mbox{Appendix~\ref{app:#1}}}
\newcommand{\ct}{\cos\theta_t}

\newcommand{\lsim}{\;\raisebox{-0.9ex}{$\textstyle\stackrel{\textstyle<}
           {\sim}$}\;}

\begin{document}


\begin{titlepage}
\begin{flushright}
HEPHY-10702
\end{flushright}

\begin{center}
{\large\bf\boldmath
    CP violating asymmetries in single top quark production
    at the Tevatron $p\bar p$ collider}\\

\vspace{1cm}
 {\bf Ekaterina Christova}\footnote{
 e-mail: echristo@inrne.bas.bg}\\
{\em Institute of Nuclear Research and Nuclear Energy, \\
 Tzarigradsko Chaussee 72, Sofia 1784, Bulgaria} \\

\vspace{0.5cm}

{\bf Stephan Fichtinger\footnote{
 e-mail: stephan@qhepu3.oeaw.ac.at}, Sabine Kraml\footnote{
 e-mail: kraml@qhepu3.oeaw.ac.at}, Walter Majerotto\footnote{
 e-mail: majer@qhepu3.oeaw.ac.at}} \\
{\em Institut f\"ur Hochenergiephysik der
 \"Osterreichischen Akademie der Wissenschaften, \\
 A-1050 Vienna, Austria}

\end{center}
\vspace{0.5cm}
\begin{abstract}

Analytic expressions for the angular distributions of the
$b$-quarks associated with single $t$-quark production in $p\bar
p \to W^* \to t\bar b \to b\bar b W$ and of the leptons from the
subsequent decay $W\to l\nu$ are obtained in the laboratory system. CP
violation in the $t$-production vertex is assumed. Different
angular and total cross section CP violating asymmetries are
considered.
Relations testing CP violation in the $t$-decay vertex are also given.
A numerical analysis is
performed in the MSSM with a CP violating phase of the trilinear
coupling $A_{\tilde t}$. The asymmetries are typically of the order
$10^{-3} - 10^{-4}$.

\end{abstract}
\end{titlepage}
\newpage
\setcounter{page}{1}


\section{Introduction \label{sect:intro}}

The experiments at the Tevatron $p\bar p$ collider offer the possibility of
thoroughly studying the properties of the top quark.
In particular, in top physics the Standard Model (SM) gives negligible
CP violating effects \cite{SM} due to the GIM mechanism.
Looking for CP violation in top quark production and/or decays is therefore
one of the best ways to probe New Physics.
In extensions of the SM as, for instance, in two Higgs doublet models or
in supersymmetric (SUSY) models, CP violating phases appear rather naturally.
They can then cause CP violating effects in processes with top quarks
at one--loop level.

In this paper we study CP violating asymmetries
in $t\bar b$ and $\bar tb$ production in $p\bar p$ collisions,
induced by CP violating form factors in the $tbW$ vertex.
More precisely, we consider the two CP conjugate processes
\begin{equation}
  p \bar{p} \to W^+ \to t \bar{b} \quad {\rm and} \quad
  p \bar{p} \to W^- \to b \bar{t} \label{eq:ppbar}
\end{equation}
followed by the decays $t \to bW^+ \to b l^+\nu_l$, $\bar{t} \to \bar bW^-
\to \bar{b} l^-\bar\nu_l$. The possibility of testing CP
violation in $p\bar p\to t\bar b\to W^+b\bar b$ was already
considered in \cite{Soni96,Soni98}  and in more details in a
review on this subject in \cite{Sonireview}. In this article, we
present new additional CP violating asymmetries.
In particular, we define
two types of asymmetries: (i) with $b$ quarks and (ii) with
leptons in the final states. For the asymmetries with $b$ quarks
we assume that the $b$'s from production can be distinguished from
those from $t$ decays.
This is justified by the requirement that
the whole event be reconstructed in order to distinguish the events of
single $t$ production in (\ref{eq:ppbar}) from the background events
\cite{background}. We thus define separate asymmetries for $b$
quarks from production, from decay, and from both production and
decay. The asymmetries with leptons are in general easier to
observe. However, the cross section is smaller by the branching
ratio $W\to l\nu$, which is roughly $1/3$ if all leptons $l=e,\mu,\tau$
are counted. \\
For both $b$ quarks and leptons, we consider asymmetries in the
total number of $b$ and $\bar b$ ($l^+$ and $l^-$) and
asymmetries in their angular distributions (forward--backward asymmetries).
When considering the angular distributions of the decay products
we assume CP violation only in the production vertex of the $t$ quark.
As shown in \cite{review} CP violation in $t$-decay is suppressed
by the polarization of the $t$-quark. Moreover,  CP violation in the
decay would mean that in addition to CP violating phases there are
also new decay modes of the top quark, which
--according to the present experimental limits--
does not seem to be the case.
However, we will also discuss the contribution of CP violation in the
$t$ decay to the total cross section asymmetries defined
in Sections 4.2 and 5.

The paper is organized as follows: In \sect{notation} we briefly
give our notation. In \sect{production} we define the CP
violating asymmetries for $b$ quarks from production, namely
total cross section asymmetries and forward--backward asymmetries.
In \sect{decay} we derive the formulae for the asymmetries for $b$
quarks and leptons from the $t$ decay.
We again consider total
cross section asymmetries and forward--backward asymmetries. The
forward--backward asymmetry for $b$ quarks from both production
and decay is given in \sect{prodec}. In \sect{numerics} we give a
numerical analysis and discussion of the resulting asymmetries in
the framework of the Minimal Supersymmetric Standard Model (MSSM)
with complex parameters. The necessary formulae for the form
factors, mass matrices, and couplings in the MSSM are given in
the Appendices.


\section{Notation \label{sect:notation}}

The quark subprocesses which we consider for single $t$ and $\bar t$
production are
\begin{eqnarray}
  u + \bar d &\to& W \:\to\:\, t + \bar b \,, \label{eq:1} \\
  d + \bar u &\to& W \:\to\:\, b + \bar t \,. \label{eq:2}
\end{eqnarray}

\noi The matrix elements including CP violation read $(m_b=0)$:
\begin{eqnarray}
  M^{t\bar b} &=& \frac{-g^2}{2}\:
    \bar u(-p_{\bar d})\,\gamma_\alpha P_L\, u(p_u)\,
    \frac{i}{\hat s-m_W^2}\,
    \bar{u}(p_t)\, \Gamma^\alpha\, u(-p_{\bar b}) \,,\\
  M^{b\bar t} &=& \frac{-g^2}{2}\:
    \bar u(-p_{\bar u})\,\gamma_\alpha P_L\, u(p_d)\,
    \frac{i}{\hat s-m_W^2}\,
    \bar{u}(p_b)\, \bar\Gamma^\alpha\, u(-p_{\bar t})
\end{eqnarray}
with
\begin{eqnarray}
  \Gamma^{\alpha} &=& \gamma^\alpha P_L (1+if_L^{CP})
     + \frac{P^\alpha}{m_t} P_L\,i\,g_R^{CP}, \\
  \bar\Gamma^{\alpha} &=& \gamma^\alpha P_L (1-if_L^{CP})
     - \frac{\bar P^\alpha}{m_t} P_L\,i\,g_R^{CP},
\end{eqnarray}
$P^\alpha = p_t^\alpha  - p_{\bar b}^\alpha$, $\bar P^\alpha =
p_b^\alpha  - p_{\bar t}^\alpha$ and $P_L=(1-\gamma_5
)/2$. $f_L^{CP}$ and $g_R^{CP}$ are the CP violating form
factors of the $tbW$ vertex. They are complex functions. The
asymmetries considered in this paper measure the absorptive parts
of these form factors --- $\Im m\,f_L^{CP}$ and $\Im m\,g_R^{CP}$.
The real part of $g_R^{CP}$, $\Re e\,g_R^{CP}$, can be measured
through triple product correlations.
As there is no CP violation at tree level, $\Re e\,f_L^{CP}$
has no physical meaning and cannot enter measurable quantities.


\section{\boldmath $b$ quarks from production \label{sect:production}}

In the centre--of--mass system (CMS) of $u$ and $\bar d$ or $\bar u$ and $d$,
with the $z$ axis pointing along the momentum $p_u$ or $p_d$,
the angular distributions of the $b$ and $\bar b$ quarks in the
subprocesses \eq{1} and \eq{2} are:
\begin{eqnarray}
  d\hat\sigma_1^{\bar tb} &=& {\cal C}
    \left\{ a_0^b + a_1^b \, \cos\theta_{b}^*
            + a_2^b \,\cos^2\theta_{b}^* \right\}\, d\cos\theta_{b}^* \,,\\
  d\hat\sigma_1^{t\bar b} &=& {\cal C}
    \left\{ a_0^{\bar b} - a_1^{\bar b}\,\cos\theta_{\bar b}^*
            + a_2^{\bar b}\,\cos^2\theta_{\bar b}^* \right\}\,
    d\cos\theta_{\bar b}^* \,.
\end{eqnarray}
Here
\begin{eqnarray}
  a_i^{b,\bar b}\;\, &=&  a_i^{SM} \pm a_i^{CP} \,, \\
  a_0^{SM}  &=& -(m_t^2 + \hat s)/2 \,,\\
  a_1^{SM}  &=& -\hat s \,,\\
  a_2^{SM}  &=& (m_t^2 -\hat s)/2 \,,\\
  a_0^{CP}\,&=& 2 \,a_0^{SM}\,\Im m\,f_L^{CP}(\hat s) +
                (\hat s -m_t^2) \, \Im m\,g_R^{CP}(\hat s)\,,\\
  a_1^{CP}\,&=& 2\, a_1^{SM}\,\Im m\,f_L^{CP}(\hat s) \,,\\
  a_2^{CP}\,&=& 2\, a_2^{SM}\left(\Im m\,f_L^{CP}(\hat s)  + \,
                \Im m\,g_R^{CP}(\hat s)\right)\,,
\end{eqnarray}
and
\begin{equation}
  {\cal C} = \frac{ -\pi\alpha_w^2(\hat s-m_t^2)^2 }{
                        16[\hat s(\hat s-m_W^2)]^2 } \,.
\end{equation}
$\hat s$ stands for
\begin{equation}
  \hat s = \left\{ \begin{array}{l}
    \hat s_- = (p_{\bar u}+p_d)^2 \quad {\rm for}
                                  \quad \hat\sigma^{\bar tb} \\
    \hat s_+ = (p_u+p_{\bar d})^2 \quad {\rm for}
                                  \quad \hat\sigma^{t\bar b}
  \end{array} \right.
\end{equation}

\noi
The polar angle $\cos\theta$ in the laboratory frame
--- the CMS of $p$ and $\bar p$ ---
is related  to the angle $\cos\theta^*$ in the CMS of the initial quarks by:
\begin{equation}
  \cos\theta^* = \frac{\cos\theta - v}{1-v\cos\theta} \,,
  \label{eq:Lorentz}
\end{equation}
where $v$ is the velocity of the laboratory system in the CMS of the quarks,
$v=(x_1-x_2)/(x_1+x_2)$.
Here $\cos\theta$ stands for $\cos\theta_b$ or $\cos\theta_{\bar b}$, and
$x_1$ and $x_2$ are the fractions of the longitudinal momenta of the quarks.
The angular distribution of the $b$ and $\bar b$ quarks in the laboratory system
thus is
$\frac{d\hat\sigma_1}{d\cos\theta} =
 \frac{d\hat\sigma_1}{d\cos\theta^*}\,
 \frac{d\cos\theta^*}{d\cos\theta^{\hphantom{*}}}$.
We hence obtain:
\begin{eqnarray}
  \frac{d\hat\sigma_1^{\bar tb}}{d\cos\theta_b}
  &=& \frac{{\cal C}(1-v^2)}{(1-v\cos\theta_b)^4} \:
      \left[\, a_0^b (1-v\cos\theta_b )^2  \right. \nn \\
  & & \hspace{30mm}\left.
          +\:a_1^b (\cos\theta_b -v)(1-v\cos\theta_b ) +
             a_2^b (\cos\theta_b -v)^2 \right] \,, \\[3mm]
  \frac{d\hat\sigma_1^{ t\bar b}}{d\cos\theta_{\bar b}}
  &=& \frac{{\cal C}(1-v^2)}{(1-v\cos\theta_b)^4} \:
      \left[\, a_0^{\bar b} (1-v\cos\theta_{\bar b} )^2 \right. \nn \\
  & & \hspace{30mm}\left.
          -\:a_1^{\bar b} (\cos\theta_{\bar b}-v)(1-v\cos\theta_{\bar b}) +
             a_2^{\bar b} (\cos\theta_{\bar b}-v)^2 \right] \,.
\end{eqnarray}

\noi
Taking into account the two possibilities
$x_1=x_u$, $x_2=x_d$ and $x_1=x_d$, $x_2=x_u$,
and using the parton distribution functions $f_u$ and $f_d$ of the proton,
we get in the laboratory system:
\begin{eqnarray}
  \frac{d\sigma_1^{\bar tb}}{d\cos\theta_b} &=& \frac{1}{2}
    \int \frac{d\hat\sigma_1^{\bar tb}(\hat s,v)}{d\cos\theta_b}\,
    \left[f_{u}(x_{u})f_{d}(x_d) + f_{u}(x_{d})f_{d}(x_u)\right]\,
    dx_u\,dx_d \,, \label{eq:dsig1a} \\[3mm]
  \frac{d\sigma_1^{t\bar b}}{d\cos\theta_{\bar b}} &=& \frac{1}{2}
    \int \frac{d\hat\sigma_1^{t\bar b}(\hat s,v)}{d\cos\theta_{\bar b}}\,
    \left[f_{u}(x_{u})f_{d}(x_d) + f_{u}(x_{d})f_{d}(x_u)\right] \,
    dx_u\,dx_d  \label{eq:dsig1b}
\end{eqnarray}
with
\begin{equation}
  \hat s =x_{u}\,x_d\, s  \quad {\rm and} \quad
   v = (x_u-x_d)/(x_u+x_d) \,.\label{shatv}
\end{equation}

\subsection{\boldmath Total cross section asymmetries for $b$ from production}

For the total cross sections of \eq{1} and \eq{2} we obtain:
\begin{eqnarray}
  (\hat\sigma_1)^{b,\bar b}
  &=& 2\,{\cal C}\left(a_0^{b,\bar b} + a_2^{b,\bar b}/3 \right) \nn \\
  &=& (\hat\sigma_1)^{SM}(\hat s)
      \left\{ 1\pm 2\,\Im m\,f_L^{CP} \pm
              \frac{m_t^2 - \hat s}{m_t^2 + 2 \hat s}\,
              2\,\Im m\,g_R^{CP}
      \right\}
  \label{eq:sig1tot}
\end{eqnarray}
where
\begin{equation}
  (\hat\sigma_1)^{SM}(\hat s) =
  \frac{ \pi\alpha_w^2(\hat s -m_t^2)^2(m_t^2 + 2\hat s) }{
         24[\hat s (\hat s - m_W^2)]^2} \,,
  \label{eq:sig1SM}
\end{equation}
and $\alpha_w = g^2/4\pi$.
In \eq{sig1tot} the upper sign stands for $\hat\sigma^b$
and the lower one for $\hat\sigma^{\bar b}$.
The CP violating asymmetry for the total number of $b$ and $\bar b$
quarks from the production processes \eq{ppbar} is thus given by
\begin{equation}
  {\cal R}_1^{tot}(b) = \frac{(\sigma_1)^b-(\sigma_1)^{\bar b}}
                      {(\sigma_1)^b + (\sigma_1)^{\bar b}} \,.
\end{equation}
From \eq{dsig1a}, \eq{dsig1b} and \eq{sig1tot} we obtain:
\begin{eqnarray}
  {\cal R}_1^{tot}(b)
  &=& \frac{\int {\cal C} \left(a_0^{CP} + a_2^{CP}/3\right)
                   f_u(x_u)f_d(x_d)\,dx_u\,dx_d }{
              \int {\cal C} \left(a_0^{SM} + a_2^{SM}/3\right)
                   f_u(x_u)f_d(x_d)\,dx_u\,dx_d } \nn\\[3mm]
  &=& \frac{2\int\hat\sigma^{SM}(\hat s)
              \left[\Im m\,f_L^{CP}(\hat s )
                    + \frac{m_t^2-\hat s}{m_t^2+2\hat s}\,
                      \Im m\,g_R^{CP}(\hat s)\right]
              f_u(x_u)f_d(x_d)\,dx_u\,dx_d }{
            \int\hat\sigma^{SM}(\hat s)\,f_u(x_u)f_d(x_d)\,dx_u\,dx_d}\,.
\label{eq:calR1}
\end{eqnarray}
This asymmetry was already obtained and discussed in
\cite{Soni98}. We recall it here for completeness and
also include it in our numerical analysis.

\subsection{\boldmath Forward--backward asymmetries for $b$ from production}

We next define two CP violating forward--backward asymmetries
for the production process:
\begin{eqnarray}
  A_1^{FB}(b) &=&
  \frac{ (\sigma_1)^b_F - (\sigma_1)^{\bar b}_B }{
         (\sigma_1)^b_F + (\sigma_1)^{\bar b}_B}    \nn \\[2mm]
  &=& \frac{
  \int\,[(\hat\sigma_1)^b_F - (\hat\sigma_1)^{\bar b}_B ]\,
        [f_{u}(x_{u})f_{d}(x_d) + f_{u}(x_{d})f_{d}(x_u)]\,
        dx_{u}\,dx_{d} }{
  \int\,[(\hat\sigma_1)^b_F + (\hat\sigma_1)^{\bar b}_B ]\,
        [f_{u}(x_{u})f_{d}(x_d) + f_{u}(x_{d})f_{d}(x_u)]\,
        dx_{u}\,dx_{d} } \label{eq:A1FB}
\end{eqnarray}
and
\begin{eqnarray}
  A_2^{FB}(b) &=&
  \frac{ (\sigma_1)^b_B - (\sigma_1)^{\bar b}_F }{
         (\sigma_1)^b_B + (\sigma_1)^{\bar b}_F}    \nn \\[2mm]
  &=& \frac{
  \int\,[(\hat\sigma_1)^b_B - (\hat\sigma_1)^{\bar b}_F ]\,
        [f_{u}(x_{u})f_{d}(x_d) + f_{u}(x_{d})f_{d}(x_u)]\,
        dx_{u}\,dx_{d} }{
  \int\,[(\hat\sigma_1)^b_B + (\hat\sigma_1)^{\bar b}_F ]\,
        [f_{u}(x_{u})f_{d}(x_d) + f_{u}(x_{d})f_{d}(x_u)]\,
        dx_{u}\,dx_{d}} \label{eq:A2FB} \,.
\end{eqnarray}

\noi
Here $(\sigma_1)^b_{F(B)}$ are the number of $b$ quarks from production
in the forward (backward) direction.
Note that in single $t$ production, contrary to $t\bar t$ pair production,
we have $\vert A_1^{FB}\vert \neq \vert A_2^{FB}\vert$.
This is due to the fact that CP violation leads to a difference
in the total cross sections of $b$ and $\bar b$ production.
In terms of form factors we obtain:

\begin{eqnarray}
  A_1^{FB}(b)  &=&
\frac{\int {\cal C} \left(a_0^{CP}+(1-v^2)\,a_1^{CP}/2+a_2^{CP}/3\right)
                 f_u(x_u)f_d(x_d)\,dx_u\,dx_d }
           {\int {\cal C} \left(a_0^{SM}+(1-v^2)\,a_1^{SM}/2+a_2^{SM}/3\right)
                 f_u(x_u)f_d(x_d)\,dx_u\,dx_d} \nn\\[3mm]
 &=&
    \frac{2\int h(\hat s)\,
    [(2m_t^2 + \hat s (7-3v^2))\,\Im m\,f_L^{CP} + 2(m_t^2-\hat s)\,
\Im m\,g_R^{CP}]\,
    f_u(x_u)f_d(x_d)\,dx_u\,dx_d  }{
    \int h(\hat s)\,
    [2m_t^2 + \hat s(7-3v^2)]\,f_u(x_u)f_d(x_d)\,dx_u\,dx_d } \,, \nn \\
  & & \\[3mm]
  A_2^{FB}(b) &=&
 \frac{\int {\cal C} \left(a_0^{CP}-(1-v^2)\,a_1^{CP}/2+a_2^{CP}/3\right)
                 f_u(x_u)f_d(x_d)\,dx_u\,dx_d }
           {\int {\cal C} \left(a_0^{SM}-(1-v^2)\,a_1^{SM}/2+a_2^{SM}/3\right)
                 f_u(x_u)f_d(x_d)\,dx_u\,dx_d} \nn\\[3mm]
 &=&
    \frac{2\int h(\hat s)\,
    [(2m_t^2 + \hat s(3v^2+1))\,\Im m\,f_L^{CP} + 2(m_t^2 -\hat s)\,
\Im m\,g_R^{CP}]\,
    f_u(x_u)f_d(x_d)\,d\,x_{u}\,dx_{d} }{
    \int h(\hat s)\,
    [2m_t^2 + \hat s(3v^2+1)]\,f_u(x_u)f_d(x_d)\,dx_u\,dx_d } \nn \\
    & &
\end{eqnarray}

with
\begin{equation}
  h(\hat s) = \left[\frac{m_t^2-\hat s}{\hat s(\hat s - m_W^2)}\right]^2 \,.
\end{equation}


\section{\boldmath Secondary $b$ quarks and leptons \label{sect:decay}}

Let us now turn to the $b$ quarks and leptons which originate from the
$t$ decays:
\begin{equation}
  t \to bW \,, \qquad t \to bl^+\nu \,.
\end{equation}

\noi
Following \cite{Kawasaki,Bilenky} in the narrow width approximation
($\Gamma_t \ll m_t$) we obtain in the CMS of $u\bar d$:

\begin{eqnarray}
  d\hat\sigma_2^{x}
  &=& \left(\frac{d\hat\sigma_1^{t}}{d\ct^*}\right)\,
      d\ct^*\:\frac{d\,\Gamma (\vec{t}\to x...)}{\Gamma_{tot}}\,
      \frac{E_{t}^*}{m_{t} } \nn\\
  &=& \left(\frac{d\hat\sigma_1^{t}}{d\ct^*}\right)\,
      d\ct^*\:\frac{d\,\Gamma}{\Gamma}(\vec{t}\to x+...)\,
      Br(t\to xX) \:,
\label{eq:gen}
\end{eqnarray}

\noi
and analogously for the decay of $\bar t$.
In \eq{gen}, $d\hat\sigma_1^{t}/d\ct^*$ is the distribution
of the $t$ quarks in \eq{1}:
\begin{equation}
   \frac{d\hat\sigma_1^{t}}{d\ct^*} = {\cal C}\left\{
   a_0^{\bar b} + a_1^{\bar b}\,\ct^* + a_2^{\bar b}\,\ct^{*2}
   \right\} \,,
\end{equation}
and $E_t^*$ is the energy of the decaying $t$.
$\frac{d\,\Gamma}{\Gamma}(\vec{t}\to x+...)$ is the angular distribution
of the secondary particle $x$ ($x=b,l^+$) in this frame with $t$
polarized, normalized to the partial decay width $\Gamma (t\to x)$.
The branching ratio $Br(t\to xX)$ stands for $Br(t\to bW^+)$ or
$Br(t\to bl^+\nu_l)$.
If $\beta_t^*$ is the velocity of $t$ in the CMS of $u\bar d$
we have \cite{Sehgal}:
\begin{equation}
  \left(\frac{d\,\Gamma}{\Gamma}\right)(\vec{t}\to x+...) =
  \frac{d\Omega_x^*}{4\pi} \frac{m_t^{2}}{E_t^{*2}\left(1-\beta_t^*
  \cos\theta_{tx}^*\right)^2}\left\{1+\alpha_x\frac{(\xi
  p_x)}{(p_tp_x)}\right\} \,,
\label{eq:dgam}
\end{equation}
where $\alpha_x$ determines the sensitivity of the particle $x=b,l$
to the polarization of the $t$ quark:
\begin{eqnarray}
  \alpha_b &=& \frac{m_t^2 - 2\,m_W^2}{m_t^2 + 2\,m_W^2} \,,\\
  \alpha_l &=& -1  \,.
\end{eqnarray}

\noi
$\xi$ is the polarization four--vector of $t$, which is
determined by the production process \eq{1}, and
$\cos\theta_{tx}^*$ is the angle between the momenta $\vec{p}_t$
and $\vec{p}_x$:
\begin{equation}
  \cos\theta_{tx}^* =
  \sin\theta_t^*\sin\theta_x^*\cos\phi_x^* +
  \cos\theta_t^*\cos\theta_x^*.
\end{equation}

\noi
The treatment of the $t$ polarization four--vector and the
general formula for the differential cross section, in the CMS of
($u\bar d$), in terms of the SM- and CP-violating components of
$\xi$, needed to derive the angular distribution
$d\hat\sigma_2^x/d\cos\theta_x$ are given in \app{polarization}.

\subsection{Angular distribution of the decay products}

Integrating \eq{general} over $d\cos\theta_t^*$ and $d\phi_x^*$ we
obtain the angular $\cos\theta_x^*$ distribution of the decay
products $x=b, l^+$ in the CMS of $u\bar d$:
\begin{equation}
  \frac{d\hat\sigma_2^x}{d\cos\theta_x^*} = {\cal B}\,
  [b_0 + b_1 \cos\theta_x^* + b_2\cos^2\theta_x^*]
\label{eq:sigmadecay}
\end{equation}
where
\begin{eqnarray}
  b_i^{\hphantom{SM}} &=& b_i^{SM} + b_i^{CP} \,, \label{eq:bi} \\
  b_i^{SM} &=& c_i^{SM} + \alpha_x\, d_i^{SM} \,,\\
  b_i^{CP} &=& c_i^{CP} + \alpha_x \,d_i^{CP} \,.
\end{eqnarray}
For the CP conserving part we obtain:
\begin{eqnarray}
  c_0^{SM} &=& \frac{1}{2m_t^2}\left[2\hat sm_t^2(m_t^2+\hat s)
               \ln\frac{m_t^2}{\hat s} - (m_t^6 + 3\hat sm_t^4 -5{\hat s}^2
               m_t^2 + {\hat s}^3)\right] \,,\\
  c_1^{SM} &=& \frac{\hat s}{m_t^2}\left[-2sm_t^2\ln\frac{m_t^2}{\hat s}
               + m_t^4 -{\hat s}^2\right] \,,\\
  c_2^{SM} &=& \frac{1}{2m_t^2}\left[-6m_t^2\hat s(m_t^2+\hat s)
               \ln\frac{m_t^2}{\hat s} + (m_t^2-\hat s)(m_t^4 +10\hat s
               m_t^2 +{\hat s}^2)\right] \,,\\
  d_0^{SM} &=& -\hat s\left[(m_t^2+\hat s) \ln\frac{m_t^2}{\hat s}
               -2(m_t^2-\hat s)\right] \,,\\
  d_1^{SM} &=& 2\hat s\left[\hat s\ln\frac{m_t^2}{\hat s}
               - (m_t^2-\hat s)\right] \,,\\
  d_2^{SM} &=& -3d_0^{SM} \label{eq:d2SM} \,.
\end{eqnarray}
For the CP violating part we have:
\begin{eqnarray}
  c_0^{CP} &=& -2\,c_0^{SM}\,\Im mf_L^{CP} + \frac{m_t^2+\hat s}{m_t^2}
               \left[-2\hat s\,m_t^2\ln\frac{m_t^2}{\hat s}
               + m_t^4 - s^2\right]\Im m\,g_R^{CP} \,,\\
  c_1^{CP} &=& -2\,c_1^{SM}\,\Im m\,f_L^{CP}\\
  c_2^{CP} &=& -2\,c_2^{SM}\,\Im m\,f_L^{CP} \nn\\
           & & -\,\frac{1}{m_t^2}\left[-6m_t^2\hat s\,(m_t^2+\hat s)
               \ln\frac{m_t^2}{\hat s} + m_t^6 + 9\hat s\,m_t^2\,(m_t^2-\hat s)
               - \hat s^3\right]\Im mg_R^{CP} \,,\\
  d_0^{CP} &=& -2\,d_0^{SM}\,\Im mf_L^{CP} \nn\\
           & & +\,\frac{\hat s}{m_t^2}\left[2\,m_t^2(m_t^2+2\hat s)
               \ln\frac{m_t^2}{\hat s} - 5m_t^4 + 4\hat s\,m_t^2
               + \hat s^2\right]\Im m\,g_R^{CP} \,,\\
  d_1^{CP} &=& -2\,d_1^{SM}\,\Im m\,f_L^{CP} + \frac{2\hat s}{m_t^2}
               \left[-2\hat s\,m_t^2\,\ln\frac{m_t^2}{\hat s}
               + m_t^4-{\hat s}^2\right]\Im m\,g_R^{CP} \,,\\
  d_2^{CP} &=& -3\,d_0^{CP} \label{eq:d2CP}
\end{eqnarray}
The coefficient ${\cal B}$ is:
\begin{equation}
  {\cal B}=\frac{-\pi\alpha_w^2 m_t^2\,Br(t\to xX)}{8[\hat s(\hat s-m_W^2)]^2}\,.
\end{equation}

\noi
Relations \eq{d2SM} and \eq{d2CP} ensure that the polarization of the $t$
does not contribute to the total number of $b$ quarks from $t$ decay.
The angular distribution in the laboratory system is obtained from \eq{sigmadecay}
by the Lorentz boost \eq{Lorentz}:
\begin{equation}
  \frac{d\sigma_2^x}{d\cos\theta_x} =
  \frac{1}{2} \int \frac{d\hat\sigma_2^x(\hat s,v)}{d\cos\theta_x}
  \left[f_{u}(x_{u})f_{d}(x_d) + f_{u}(x_{d})f_{d}(x_u)\right]
  dx_u\,dx_d  \label{eq:sigma2ls}
\end{equation}
with
\begin{eqnarray}
  \frac{d\hat\sigma_2^x}{d\cos\theta_x}
  &=& \frac{{\cal B}(1-v^2)}{(1-v\cos\theta_x)^4}
      \left[b_0(1-v\cos\theta_x)^2 \right. \nn\\
  & & \hspace{30mm} +\,b_1 (\cos\theta_x-v)(1-v\cos\theta_x)
      \left.        +  b_2 (\cos\theta_x-v)^2 \right] \,,
  \label{eq:decayprod}
\end{eqnarray}
and $\hat s$ and $v$ defined in (\ref{shatv}).
The angular distribution of the decay products from $\bar t$ decay
is obtained from \eq{decayprod} by CP conjugation.

\subsection{Total cross section asymmetries for the decay products}

Analogously to \eq{calR1} we define a total cross section asymmetry
${\cal R}_2^{tot}(x)$ for the secondary particles $x$:
\begin{equation}
  {\cal R}_2^{tot}(x) =
  \frac{\sigma_2^{x}-\sigma_2^{\bar x}}{\sigma_2^{x}+\sigma_2^{\bar x}}\,,\quad x=b,l^+\,.
\end{equation}
This asymmetry was first suggested for $b$ quarks in \cite{Soni98}, but
without giving an explicit analytic expression.
 Using our result \eq{sigma2ls} and \eq{decayprod} we obtain:
\begin{eqnarray}
  {\cal R}_2^{tot}(x)
  &=& \frac{\int {\cal B}(b_0^{CP} + b_2^{CP}/3)
                 f_{u}(x_{u})f_{d}(x_d)\,dx_{u}\,dx_{d} }{
            \int {\cal B}(b_0^{SM} + b_2^{SM}/3)
                 f_{u}(x_{u})f_{d}(x_d)\,dx_{u}\,dx_{d}} \nn\\
  &=& \frac{\int {\cal B}(c_0^{CP} + c_2^{CP}/3)
                 f_{u}(x_{u})f_{d}(x_d)\,dx_{u}\,dx_{d} }{
            \int {\cal B}(c_0^{SM} + c_2^{SM}/3)
                 f_{u}(x_{u})f_{d}(x_d)\,dx_{u}\,dx_{d} } \,.
\label{eq:calR2}
\end{eqnarray}

\noi As there is no dependence on the $t$ polarization, \eq{calR2}
implies that the total cross section asymmetries for secondary $b$
quarks and leptons are equal in magnitude:
\begin{equation}
  {\cal R}_2^{tot}(b) ={\cal R}_2^{tot}(l).\label{calRbl}
\end{equation}
This is a consequence of CP invariance in the decay $W\to l\nu$.
Eq. \eq{calR2} also implies that
\begin{equation}
  {\cal R}_2^{tot}(x) =-{\cal R}_1^{tot}(b),\qquad x=b,l^+ \,.
  \label{eq:R12}
\end{equation}
However, \eq{R12} is not valid in general.
As already pointed out in \cite{Soni98}, ${\cal R}_2^{tot}(x)$ receives
contributions from both the production and decay vertices.
In the general case, \eq{R12} reads:
\begin{equation}
  {\cal R}_2^{tot}(x) = -{\cal R}_1^{tot}(b) +\delta_1^{CP},
  \qquad x=b,l^+ \,.
  \label{eq:R12decay1}
\end{equation}
where $\delta_1^{CP}$ is due to CP violation
in the $t$-decay~\cite{CF}
\begin{equation}
  \delta_1^{CP} =
  \frac{\Gamma (t\to bW^+) - \Gamma (\bar t\to \bar b W^-)}
       {\Gamma (t\to bW^+) + \Gamma (\bar t\to \bar b W^-)}
  \,\cdot\label{eq:deltaCP}
\end{equation}
Testing \eq{R12decay1} would be a model independent test of CP
violation in the $t$-decay vertex.

\subsection{Forward--backward asymmetries for the secondary products}

For the secondary products $x$ we can again define
two forward--backward asymmetries in the laboratory frame:
\begin{eqnarray}
  R_1^{FB}(x)
  &=& \frac{(\sigma_2)_F^x-(\sigma_2)_B^{\bar x} }{
            (\sigma_2)_F^x+(\sigma_2)_B^{\bar x} } =
      \frac{\int \left[(\hat\sigma_2)^x_F-(\hat\sigma_2)^{\bar x}_B\right]
                 f_u(x_u)f_d(x_d)\,dx_u\,dx_d }{
            \int \left[(\hat\sigma_2)^x_F+(\hat\sigma_2)^{\bar x}_B\right]
                 f_u(x_u)f_d(x_d)\,dx_u\,dx_d } \,,
      \label{eq:R1FB} \\[3mm]
  R_2^{FB}(x)
  &=& \frac{(\sigma_2)_B^x-(\sigma_2)_F^{\bar x} }{
            (\sigma_2)_B^x+(\sigma_2)_F^{\bar x} } =
      \frac{\int \left[(\hat\sigma_2)^x_B-(\hat\sigma_2)^{\bar x}_F\right]
                 f_u(x_u)f_d(x_d)\,dx_u\,dx_d }{
            \int \left[(\hat\sigma_2)^x_B+(\hat\sigma_2)^{\bar x}_F\right]
                 f_u(x_u)f_d(x_d)\,dx_u\,dx_d } \,.
      \label{eq:R2FB}
\end{eqnarray}

\noi
From \eq{sigmadecay} and \eq{bi} we obtain:
\begin{eqnarray}
  R_1^{FB}(x)
  &=& \frac{\int {\cal B} \left(c_0^{CP}+c_2^{CP}/3-(v^2-1)b_1^{CP}/2\right)
                 f_u(x_u)f_d(x_d)\,dx_u\,dx_d }
           {\int {\cal B} \left(c_0^{SM}+c_2^{SM}/3-(v^2-1)b_1^{SM}/2\right)
                 f_u(x_u)f_d(x_d)\,dx_u\,dx_d} \nn\\[3mm]
  &=& \frac{-2\int D(\hat s)
              \left\{k_1(\hat s)\,\Im m\,f_L + l_1(\hat s)\,\Im m\,g_R\right\}
              f_u(x_u)f_d(x_d)\,dx_u\,dx_d }
           {\int D(\hat s)\,k_1(\hat s)\,
                 f_u(x_u)f_d(x_d)\,dx_u\,dx_d }
\end{eqnarray}
with
\begin{eqnarray}
  k_1(\hat s)
    &=& (m_t^2-\hat s)\left[2m_t^4 -\hat s^2(7-3v^2)
        - \hat s m_t^2 ((1-3v^2)-6\,\alpha_x(1-v^2))\right] \nn\\
    & & +\,6\,(1-\alpha_x)(1-v^2)\,m_t^2\,\hat s^2\ln\frac{m_t^2}{\hat s}\,,\\
  l_1(\hat s)
    &=& (m_t^2-\hat s)\left[2(m_t^2-\hat s)^2 + 3\,\alpha_x\hat s
        (1-v^2)\,(m_t^2 +\hat s)\right] \nn\\
    & & -\,6\,\alpha_x(1-v^2)\,m_t^2\,\hat s^2\ln\frac{m_t^2}{\hat s} \,,
\end{eqnarray}
and
\begin{eqnarray}
  R_2^{FB}(x)
  &=& \frac{\int{\cal B}\left(c_0^{CP}+c_2^{CP}/3+(v^2-1)b_1^{CP}/2\right)
                 f_u(x_u)f_d(x_d)\,dx_u\,dx_d }
           {\int{\cal B}\left(c_0^{SM}+c_2^{SM}/3 +(v^2-1)b_1^{SM}/2\right)
                 f_u(x_u)f_d(x_d)\,dx_u\,dx_d } \nn\\[3mm]
  &=& \frac{-2\int D(\hat s)
              \{k_2(\hat s)\,\Im m\,f_L + l_2(\hat s)\,\Im m\,g_R\}
              f_u(x_u)f_d(x_d)\,dx_u\,dx_d }
           {\int D(\hat s)\,k_2(\hat s)\,f_u(x_u)f_d(x_d)\,dx_u\,dx_d}
\end{eqnarray}
with
\begin{eqnarray}
  k_2(\hat s)
    &=& (m_t^2-\hat s)\left[2m_t^4 -\hat s^2(1+3v^2)
        + \hat s m_t^2((5-3v^2)-6\alpha_x(1-v^2))\right] \nn\\
    & & -\,6\,(1-\alpha_x)(1-v^2)\,m_t^2\,\hat s^2 \ln\frac{m_t^2}{\hat s}\,,\\
  l_2(\hat s)
    &=& (m_t^2-\hat s)\left[2(m_t^2-\hat s)^2 - 3\alpha_x(1-v^2)\,
        \hat s\,(m_t^2 +\hat s)\right] \nn\\
    & & +\,6\,\alpha_x(1-v^2)\,m_t^2\,\hat s^2\ln\frac{m_t^2}{\hat s} \,.
\end{eqnarray}
The factor $D(\hat s)$ is
\begin{equation}
  D(\hat s) = [\hat s(\hat s-m_W^2)]^{-2} \:.
\end{equation}

\noi
Note that $R_{1,2}^{FB}$ are polarization asymmetries, i.e. they measure
different combinations of the CP violating contributions to the
$t$ quark polarization. This can also be seen from the explicit expressions
for $k_{1,2}$ and $l_{1,2}$. As a consequence, the forward--backward asymmetries
for the $b$ quarks are different from those for the leptons.
Note, moreover, that the contribution from $t$ polarization enters only
through the term $b_1$ which is linear in $\cos\theta_x$, which ensures
that  polarization does not contribute to the total cross section.


\section{\boldmath $b$ quark forward--backward asymmetry for the
                   sum of the cross sections \label{sect:prodec}}

An asymmetry that seems most convenient what concerns statistics is
the forward--backward asymmetry for the total number of $b$ and
$\bar b$ quarks from both production and decay:
\begin{equation}
  A^{FB}(b) = \frac{\sigma^b_F -\sigma^{\bar b}_B}
                {\sigma^b_F +\sigma^{\bar b}_B}   \label{eq:AFB}
\end{equation}
where
\begin{equation}
  \sigma^{b,\bar b}_F = (\sigma_1)_F^{b,\bar b} + (\sigma_2)_F^{b,\bar b}\,,
  \qquad
  \sigma^{b,\bar b}_B = (\sigma_1)_B^{b,\bar b} +(\sigma_2)_B^{b,\bar b} \,.
\end{equation}
We have:
\begin{equation}
  A^{FB}(b)=
  \frac{\int\left[{\cal C}\,a_1^{CP} + {\cal B}\,b_1^{CP}\right]
             (1-v^2)\,f_u(x_u)f_d(x_d)\,dx_u\,dx_d }
       {\int\left[4\,{\cal C}\,(a_0^{SM}+ a_2^{SM}/3) + (1-v^2)\,
                  ({\cal C}\, a_1^{SM} + {\cal B}\,b_1^{SM})\right]
            \,f_u(x_u)f_d(x_d)\,dx_u\,dx_d} \,,
\end{equation}
where we have used the fact that in the SM the total numbers of
$b$ quarks from production and decay are the same:
\begin{equation}
  ({\cal C}\,a_0^{SM} + {\cal B}\,b_0^{SM}) + ({\cal C}\,a_2^{SM}
  + {\cal B}\,b_2^{SM})/3 =2 {\cal C}(a_0^{SM}+ a_2^{SM}/3) \,.
\end{equation}

\noi
Explicitly, in terms of the CP violating form factors, we obtain:
\begin{equation}
  A^{FB}(b) =
  \frac{6\int D(\hat s)
        \left[k(\hat s)\,\Im m\,f_L^{CP}+l(\hat s)\,\Im m\,g_R^{CP}\right]
        \,f_u(x_u)f_d(x_d)\,dx_u\,dx_d }
       {\int D(\hat s)\,m(\hat s)\,f_u(x_u)\,f_d(x_d)dx_u\,dx_d}
\end{equation}
with
\begin{eqnarray}
  k(\hat s) &=& 2\hat s
    \left[(\hat s -m_t^2)(\hat s + (2-3\alpha_b)m_t^2)
          +3(1-\alpha_b)m_t^2 \hat s\ln\frac{m_t^2}{\hat s}\right] \,,\\
  l(\hat s) &=& 3 \alpha_b\hat s
    \left[m_t^4-\hat s^2 -2m_t^2\hat s\ln\frac{m_t^2}{\hat s}\right]\,,\\
  m(\hat s) &=& -2
    \left[(m_t^2-\hat s)
          \left( 2m_t^4 + m_t^2\hat s(2+3\alpha_b(1-v^2))
                 - \hat s^2(7-3v^2) \right) \right. \nn\\
    & & \left. \hspace{55mm}
    +\,3m_t^2\hat s^2(1-\alpha_b)(1-v^2)\ln\frac{m_t^2}{\hat s}\right]\,.
\end{eqnarray}

Assuming CP violation only in the $t$-production vertex,
the total cross sections  $\sigma^b$ and $\sigma^{\bar b}$
for the $b$ and $\bar b$ quarks  from both
production and decay are equal, i.e. the total cross section asymmetry
$ {\cal R}^{tot}(b)$ is zero:
\begin{equation}
  {\cal R}^{tot}(b) = \frac{\sigma^b -\sigma^{\bar b}}
                    {\sigma^b + \sigma^{\bar b}}=0\,\cdot
\label{eq:asymmSoni}
\end{equation}
However, when CP violation in $t$-decay is also considered, we have:
\begin{equation}
  {\cal R}^{tot}(b) =\frac{\delta_1^{CP}}{2}\,\cdot
\label{eq:asymmSoni1}
\end{equation}

\noi
In the general case, we thus have three different total cross section
asymmetries and they measure three different quantities.
The total cross section
asymmetry of the $b$-quarks from production, $ {\cal R}^{tot}_1(b)$,
measures CP violation in the production vertex,  the total cross section
asymmetry of the $t$-decay products, $ {\cal R}^{tot}_2(x)$, $x=b,l$
measures CP violation both in the production and decay vertices,
the total cross section
asymmetry of the $b$-quarks from both production and decay,
$ {\cal R}^{tot}(b)$,
measures CP violation in the $t$-decay vertex.

Neglecting CP violation in $t$-decay, we are left
with only one forward--backward asymmetry $A^{FB}$
when {\it all} $b$ and $\bar b$ quarks from single $t$-quark production
are counted.


\section{Numerical results and discussion within the MSSM
         \label{sect:numerics}}

In this Section we present numerical results for the discussed asymmetries
in the framework of the Minimal Supersymmetric Standard Model (MSSM) with
CP violating phases.
In the MSSM one has the gaugino mass parameters $M_1=M'$, $M_2=M$,
$M_3=m_{\tilde g}$ corresponding to the groups U(1), SU(2), SU(3), where
$m_{\tilde g}$ is the gluino mass.
We assume the GUT relations:
\begin{equation}
  m_{\tilde g} = (\alpha_s /\alpha_2) M \sim 3M\,,
  \qquad M'=(5/3) \tan^2 \theta_W M  \sim \sfrac{1}{2}\,M
\end{equation}
with a common phase of $M_i$ which can be rotated away by an $R$ rotation.
The only complex parameters relevant for our discussion thus are:
the Higgsino mass parameter $\mu = |\mu|\,e^{i\phi_\mu}$ and the SUSY
breaking trilinear couplings of the stops and sbottoms,
$A_t = |A_t|\,e^{i\phi_t}$ and $A_b = |A_b|\,e^{i\phi_b}$.
As we work in the limit of $m_b=0$ the phase $\phi_b$
does not play any role.
The phase of $\mu$ is strongly constrained by the upper bounds of
the electric dipole moments of the electron and neutron.
$\phi_\mu$ must be very small except the SUSY masses are large
($>1$~TeV) \cite{Nath} or there are strong cancellations
between the different contributions \cite{Ibrahim}.
We therefore take $\phi_\mu=0$.
After this the only relevant phase we are left
with for the considered CP violating asymmetries is the phase $\phi_t$.

In the numerical analysis we assume CP violation only in the production
vertex and not in the decay of the $t$ quark: CP violation
in the decay would mean that in addition to CP violating phases
there are also new decay modes of the top quark.
According to the present experimental limits there is only a small window
left for the decay $t\to \tilde\chi^0_1 +\tilde t_1^{}$.
A detailed analysis
of CP violation in $t$-decay was performed in~\cite{Soni98}.

\begin{figure}[p]
\begin{center}
\begin{tabular}{c@{\hspace{2cm}}c}
\epsfig{file=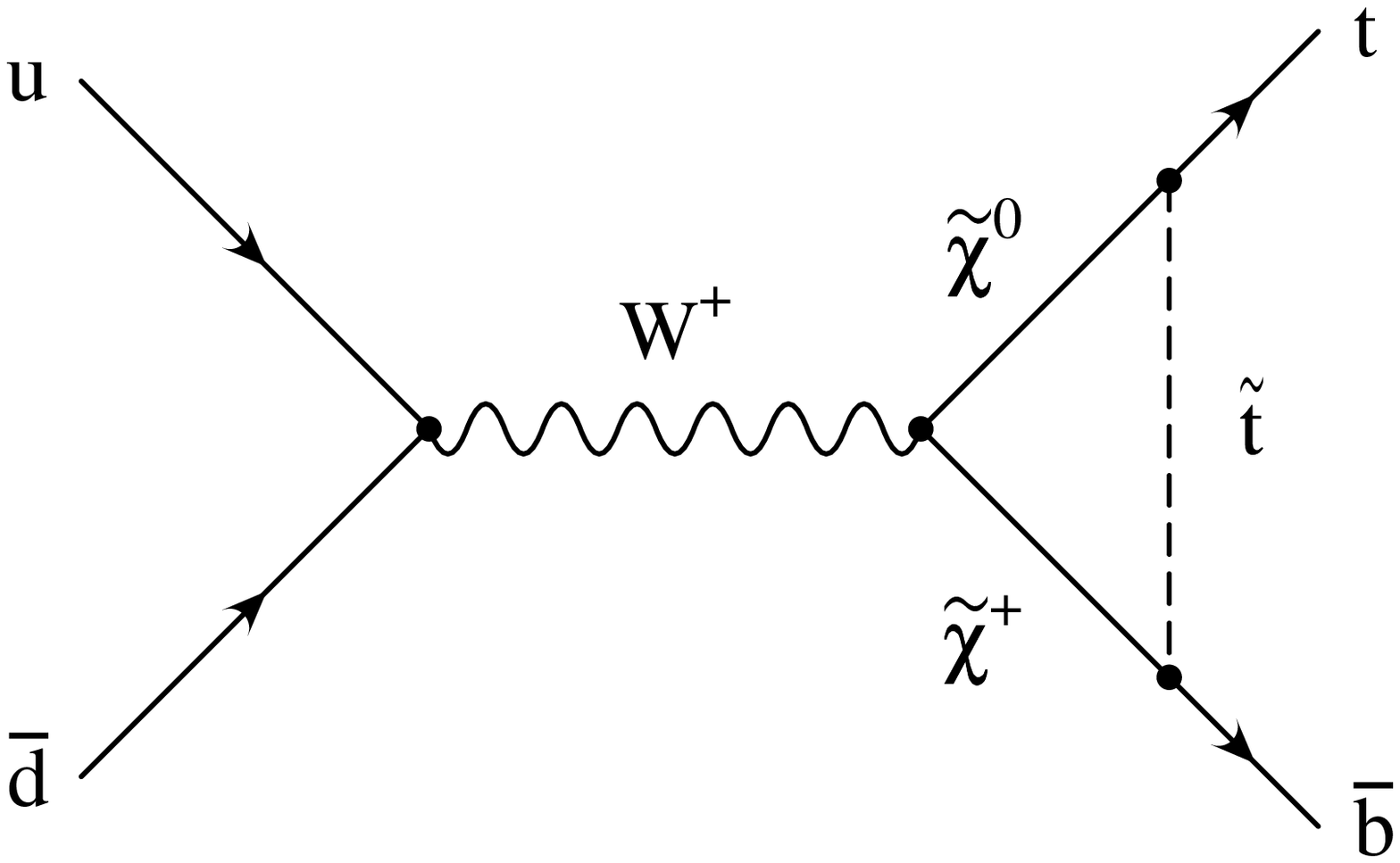,height=30mm}&
\epsfig{file=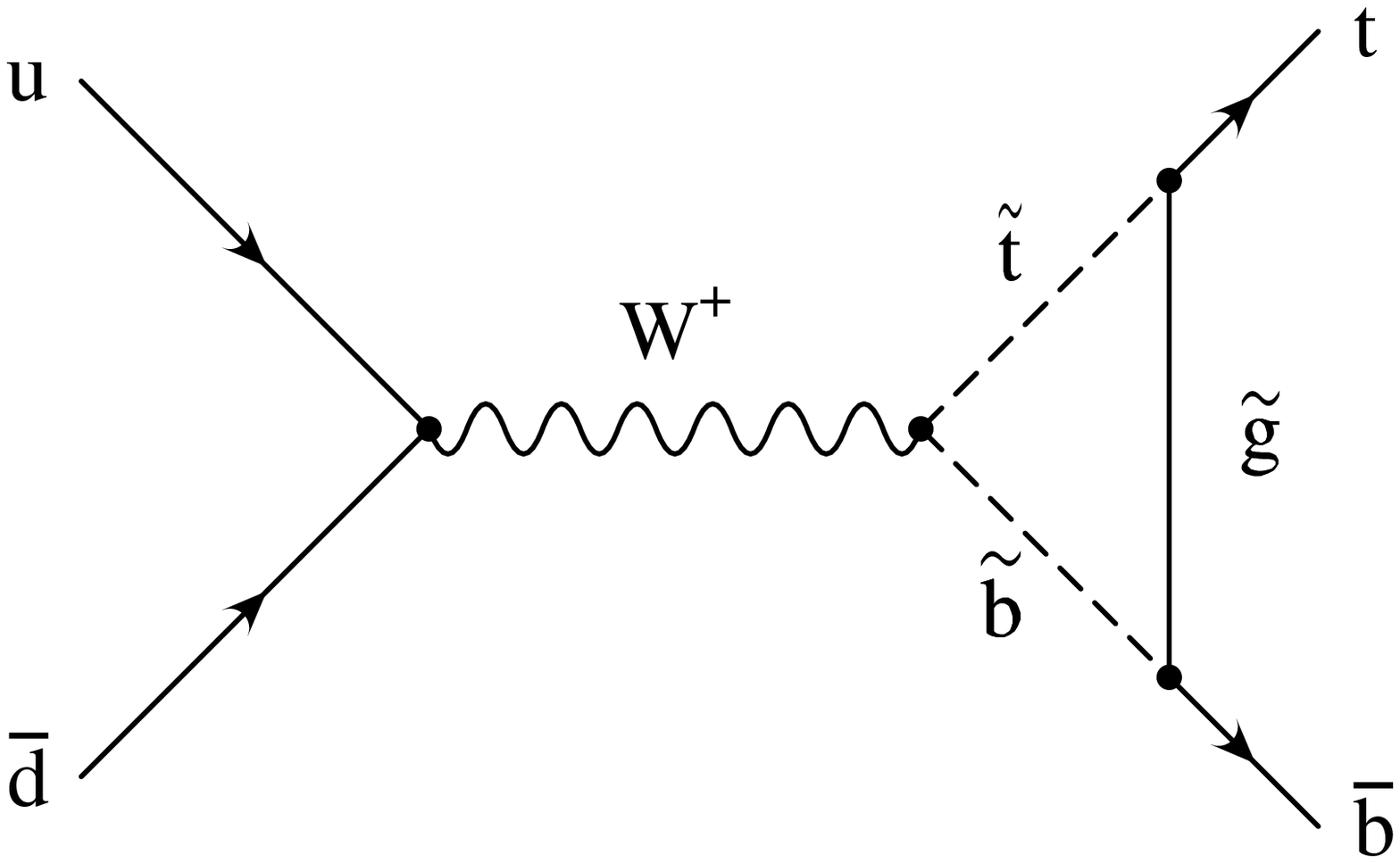,height=30mm}
\end{tabular}
\end{center}
\caption{One--loop Feynman diagrams for $u\bar d\to t\bar b$.}
\label{fig:feyndiag}
\end{figure}

At one--loop level, the reactions $u\bar b\to t\bar b$, $d\bar u\to
b\bar t$ receive radiative corrections from triangle and box
graphs with charginos $\tilde\chi^+_i$, neutralinos
$\tilde\chi^0_j$, squarks $\tilde q_i$ ($i=1,2;\;j=1...4$), and
gluinos $\tilde g$ in the loops. The analytic expressions for the
form factors due to these diagrams have been worked out in
\cite{Soni98}. Following \cite{Soni98}, in the limit $\phi_\mu \to
0$, $m_{u,d}\to 0$, only the graphs with
$(\tilde\chi^+\tilde\chi^0\,\tilde t)$ and $(\tilde t\,\tilde
b\,\tilde g)$ loops (see \fig{feyndiag}) contribute to CP
violation  and the contribution from the box diagrams is negligible.
We thus base our numerical analysis on the two contributions -- from
the $(\tilde\chi^+\tilde\chi^0\,\tilde t)$ and $(\tilde t\,\tilde
b\,\tilde g)$ loops.
Our formulae for the form factors agree with those in
\cite{Soni98} and are given in \app{formfac}.

We now want to analyze the influence of a possibly large phase of $A_t$
on the CP violating asymmetries defined in the previous Sections.
For this purpose we choose three scenarios of gaugino--higgsino mixing:
(i) a gaugino scenario with $M=116$~GeV, $\mu=400$~GeV,
(ii) a Higgsino scenario with $M=400$~GeV, $\mu=116$~GeV, and
(iii) a 'mixed' scenario with $M=\mu=168$~GeV.
Further, we take $\tan\beta=4$,
$m_{\tilde Q}=300$~GeV, $m_{\tilde U}=270$~GeV, and $m_{\tilde D}=330$~GeV.
The SM parameters are: $m_t=174$~GeV, $m_W=80.03$~GeV, $\sin^2\theta_W=0.23$
$\alpha(m_Z)=1/129$, and $\alpha_s(m_Z)=0.12$ (we neglect the bottom
mass, $m_b=0$). For the structure functions of the proton we use
the CTEQ5 parton distribution function cteq5m1 \cite{cteq}.
We leave out the theoretical uncertainties associated with
the QCD corrections and parton distribution functions
discussed in \cite{Yuan} since these uncertainties, being CP-even,
cannot mimic the CP-violating asymmetry discussed here.

Figures~\ref{fig:scen1}\,--\,\ref{fig:scen3}
show the resulting CP violating asymmetries
for the three scenarii as a function of $\phi_t$.
$|A_t|$ is chosen such that $m_{\tilde t_1}\simeq 96$~GeV.
Moreover, we have $m_{\tilde\chi^+_1}\simeq 104$~GeV and
$m_{\tilde\chi^0_1}\sim$~50--100~GeV.
The asymmetries are typically of the order of $10^{-4}$.
Note, however, that the masses in our scenarii are just at the
borderline of the experimentally allowed values \cite{lepc}.
If the mass spectrum becomes heavier the CP violating asymmetries
quickly decrease.
The asymmetries also decrease with increasing $\tan\beta$.

The asymmetries $A_2^{FB}(b)$ and $R_2^{FB}(b)$ seem to be
the most promising ones --- they reach up to $\sim 0.1\%$.
Our results are an order of magnitude smaller then the
estimates obtained in \cite{Soni98}.
A general result for the two types of forward-backward
asymmetries is that
\begin{equation}
  A_2^{FB} \geq {\cal R}_1^{tot} \geq A_1^{FB}, \qquad \qquad
  R_2^{FB} \geq R_1^{FB},
\end{equation}
and $A_2^{FB}$ ($R_2^{FB}$) in some cases
can be $\sim 2 - 3$ times bigger than $A_1^{FB}$  ($R_1^{FB}$).
Another general feature is that the
forward-backward asymmetries for the decay products
$R^{FB}_{1(2)}$ are almost equal for $b$-quarks and leptons:
\begin{equation}
  R^{FB}_{1(2)}\,(b)\simeq R^{FB}_{1(2)}\,(l)\,.
\end{equation}
The forward-backward asymmetry $A^{FB}$ for the sum of $b$ quarks
from both production and decay turns out to be extremely small
$\lsim 0.02\%$.


\section{Conclusions}

We have studied the process of single $t$-quark production in
$p\bar p$  collisions. Assuming CP violation in the $t$-production
vertex we have defined different angular and total-cross-section
asymmetries as measurable quantities.
General analytic expressions for these asymmetries in terms of
the corresponding CP violating form factors are obtained.
Relations sensitive to CP violation in the $t$-decay vertex are
defined.
We want to emphasize that the formulae are valid independently of the
origin of CP violation.

We have performed a numerical analysis within the MSSM with complex
phases.
In accordance with the constraints from the
measurements of the electric dipole moments of the electron and
the neutron we have only kept the influence of the phase of
$A_{\tilde t}$.
The effects turn out to be rather small.
The discussed asymmetries are of the order $10^{-3}$ -- $10^{-4}$
and they decrease as the mass spectrum and $\tan\beta$ increase.

The cross section for $p\bar p \to W^+ \to t\bar b$ is $\sim$~340~$fb$.
This means $\sim 10^4$ events for RUN II of the Tevatron with an
integrated luminosity of $30\,fb^{-1}$ ($15\,fb^{-1}$ per experiment).
To measure an asymmetry of the order of $10^{-4}$, one would need
 about $10^8$ events. This is far beyond
the observability of such an asymmetry at the Tevatron.
Nevertheless, it would be worthwhile to look for CP violation in top
physics as it would imply not only physics beyond the Standard Model, but
also beyond the Minimal Supersymmetric Standard Model.

\section{Acknowledgements}

This work was supported by the ``Fonds zur F\"orderung der
wissenschaftlichen Forschung'' of Austria, project no. P13139-PHY,
the EU TMR Network Contract HPRN-CT-2000-00149 and the Bulgarian
National Science Foundation, Grant PH-1010. E. Ch. is grateful to the
hospitality of the Institute of High Energy Physics in Vienna,
where this work was mostly carried out supported by the Austrian
Academy of Sciences.
\setlength{\unitlength}{1mm}

\begin{figure}[p]
\begin{center}
\begin{picture}(80,185)
\put(0,125){\mbox{\epsfig{file=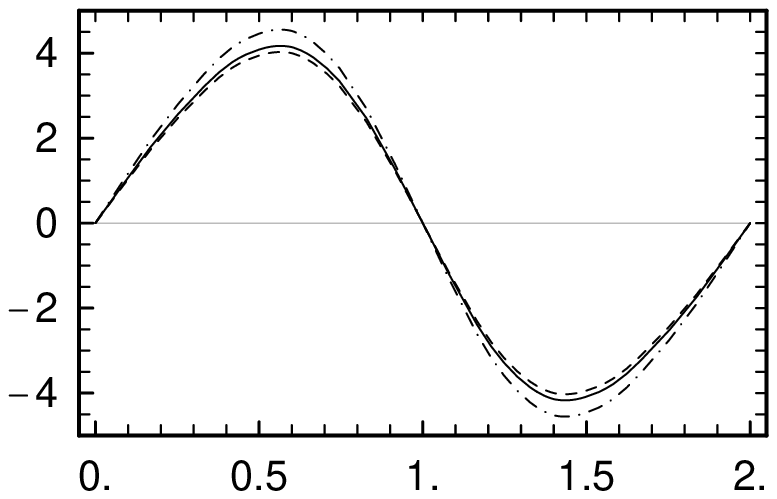,height=45mm}}}
\put(0,65){\mbox{\epsfig{file=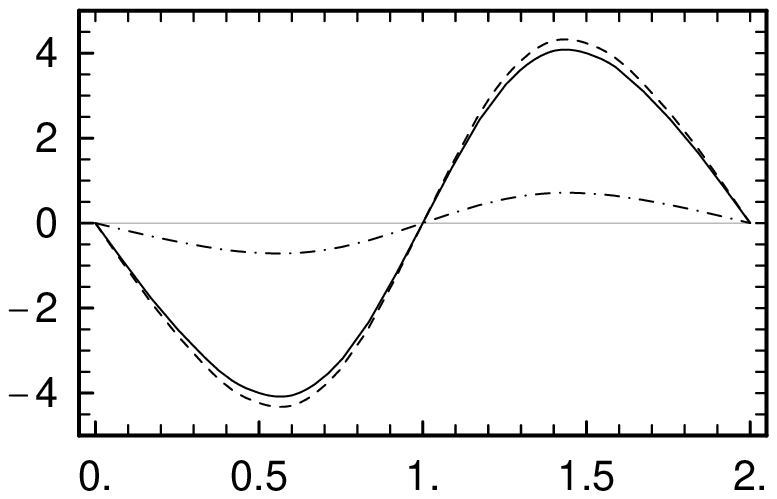,height=45mm}}}
\put(0,5){\mbox{\epsfig{file=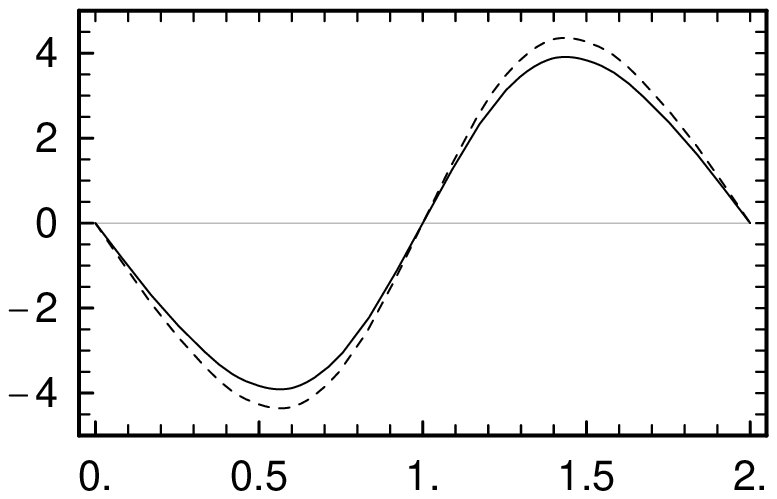,height=45mm}}}
\put(35,0){\mbox{$\phi_t~[\,\pi\,]$}}
\put(35,60){\mbox{$\phi_t~[\,\pi\,]$}}
\put(35,120){\mbox{$\phi_t~[\,\pi\,]$}}
\put(-2,13){\makebox(0,0)[br]{{\rotatebox{90}{Asymmetry $[10^{-4}]$}}}}
\put(-2,73){\makebox(0,0)[br]{{\rotatebox{90}{Asymmetry $[10^{-4}]$}}}}
\put(-2,133){\makebox(0,0)[br]{{\rotatebox{90}{Asymmetry $[10^{-4}]$}}}}
\put(48.5,142){\mbox{\scriptsize${\cal R}_1^{tot}(b)$}}
\put(53,140.5){\mbox{$\vector(0,-1){6}$}}
\put(21,159.5){\mbox{\scriptsize$A_1^{FB}(b)$}}
\put(34,164){\mbox{\scriptsize$A_2^{FB}(b)$}}
\put(21,79){\mbox{\scriptsize$R_1^{FB}(b)$}}
\put(35,76){\mbox{\scriptsize$R_2^{FB}(b)$}}
\put(49,95.5){\mbox{\scriptsize$A^{FB}(b)$}}
\put(21,19){\mbox{\scriptsize$R_1^{FB}(l)$}}
\put(34,43){\mbox{\scriptsize$R_2^{FB}(l)$}}
\end{picture}
\end{center}
\caption{CP violating asymmetries (in units of $10^{-4}$)
  for the gaugino scenario: $M=116$~GeV, $\mu=400$~GeV, $\tan\beta=4$,
  $m_{\tilde Q}=300$~GeV, $m_{\tilde U}=270$~GeV, $m_{\tilde D}=330$~GeV,
  $m_{\tilde\chi^+_1}\simeq 104$~GeV, and $m_{\tilde t_1}\simeq 96$~GeV.}
\label{fig:scen1}
\end{figure}


\begin{figure}[p]
\begin{center}
\begin{picture}(80,185)
\put(0,125){\mbox{\epsfig{file=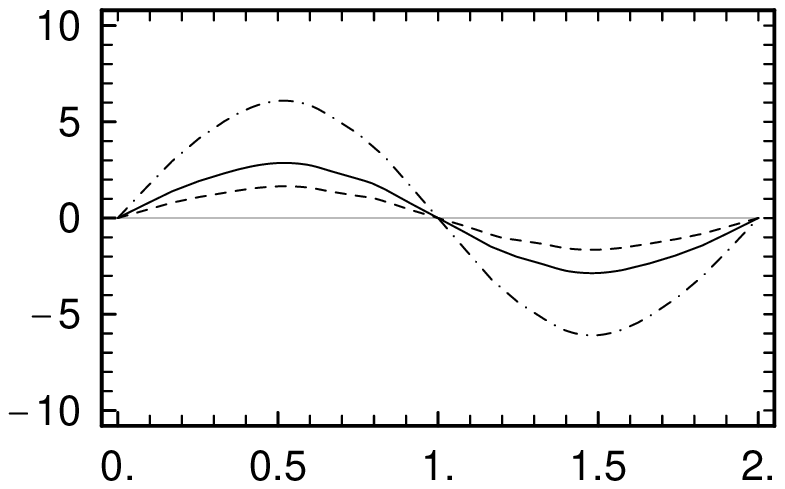,height=45mm}}}
\put(0,65){\mbox{\epsfig{file=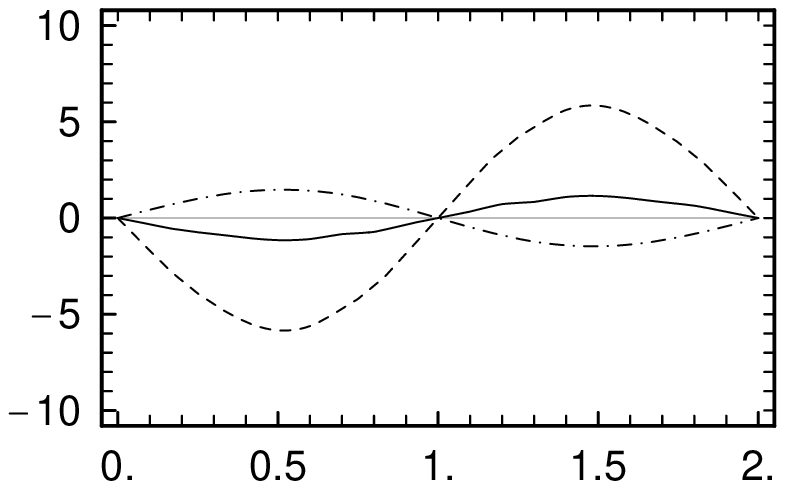,height=45mm}}}
\put(0,5){\mbox{\epsfig{file=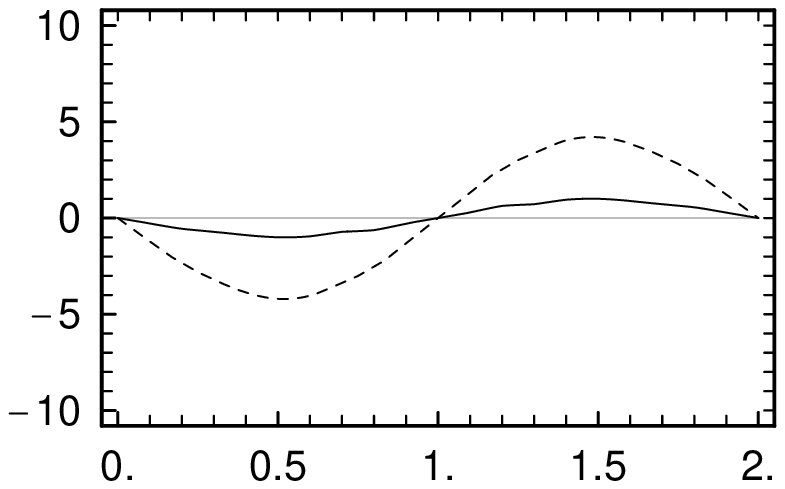,height=45mm}}}
\put(35,0){\mbox{$\phi_t~[\,\pi\,]$}}
\put(35,60){\mbox{$\phi_t~[\,\pi\,]$}}
\put(35,120){\mbox{$\phi_t~[\,\pi\,]$}}
\put(-2,13){\makebox(0,0)[br]{{\rotatebox{90}{Asymmetry $[10^{-4}]$}}}}
\put(-2,73){\makebox(0,0)[br]{{\rotatebox{90}{Asymmetry $[10^{-4}]$}}}}
\put(-2,133){\makebox(0,0)[br]{{\rotatebox{90}{Asymmetry $[10^{-4}]$}}}}
\put(37,160){\mbox{\scriptsize${\cal R}_1^{tot}(b)$}}
\put(36,159.5){\mbox{$\vector(-3,-2){6}$}}
\put(21,143){\mbox{\scriptsize$A_1^{FB}(b)$}}
\put(24,147){\mbox{$\vector(0,1){5}$}}
\put(38,137.5){\mbox{\scriptsize$A_2^{FB}(b)$}}
\put(21,84.5){\mbox{\scriptsize$R_1^{FB}(b)$}}
\put(37,100){\mbox{\scriptsize$R_2^{FB}(b)$}}
\put(50,82.5){\mbox{\scriptsize$A^{FB}(b)$}}
\put(21,35.5){\mbox{\scriptsize$R_1^{FB}(l)$}}
\put(25,33.5){\mbox{$\vector(0,-1){5}$}}
\put(40,39){\mbox{\scriptsize$R_2^{FB}(l)$}}
\end{picture}
\end{center}
\caption{CP violating asymmetries (in units of $10^{-4}$)
  for the higgsino scenario: $M=400$~GeV, $\mu=116$~GeV, $\tan\beta=4$,
  $m_{\tilde Q}=300$~GeV, $m_{\tilde U}=270$~GeV, $m_{\tilde D}=330$~GeV,
  $m_{\tilde\chi^+_1}\simeq 104$~GeV, and $m_{\tilde t_1}\simeq 96$~GeV.}
\label{fig:scen2}
\end{figure}


\begin{figure}[p]
\begin{center}
\begin{picture}(80,185)
\put(0,125){\mbox{\epsfig{file=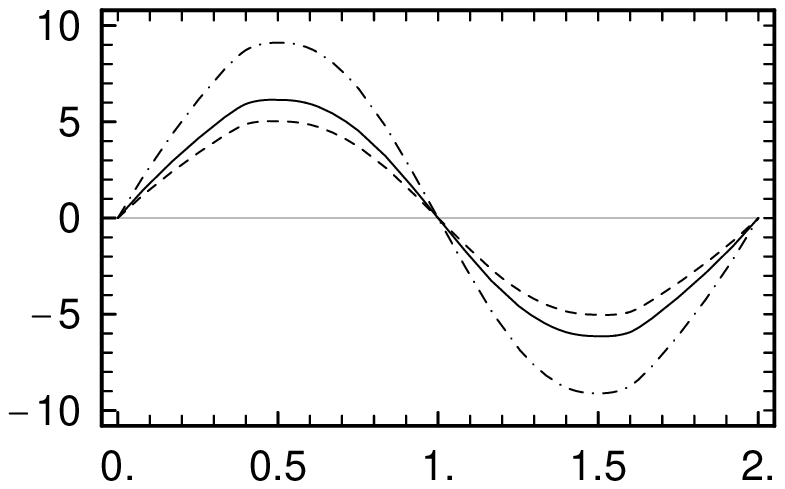,height=45mm}}}
\put(0,65){\mbox{\epsfig{file=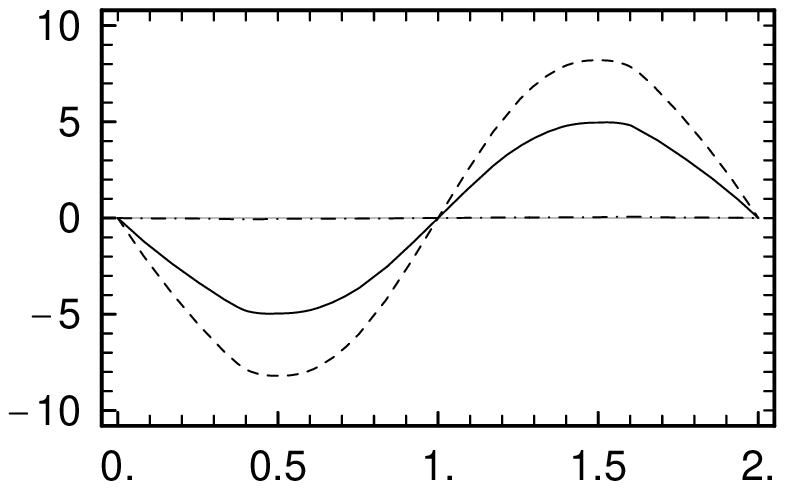,height=45mm}}}
\put(0,5){\mbox{\epsfig{file=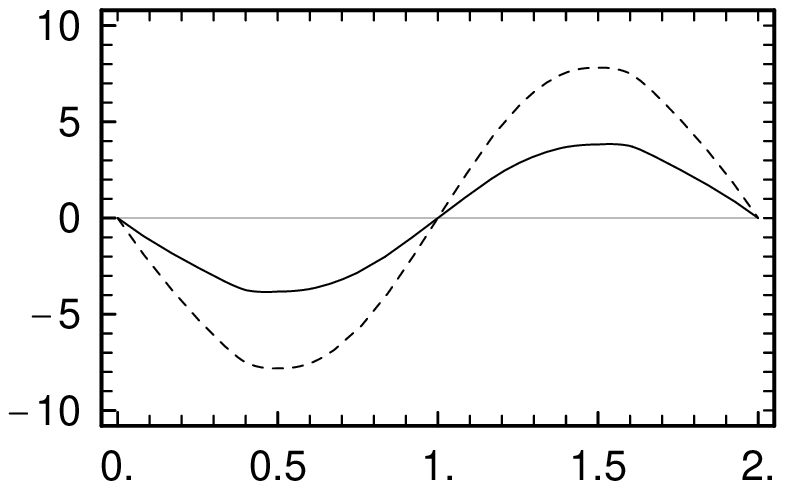,height=45mm}}}
\put(35,0){\mbox{$\phi_t~[\,\pi\,]$}}
\put(35,60){\mbox{$\phi_t~[\,\pi\,]$}}
\put(35,120){\mbox{$\phi_t~[\,\pi\,]$}}
\put(-2,13){\makebox(0,0)[br]{{\rotatebox{90}{Asymmetry $[10^{-4}]$}}}}
\put(-2,73){\makebox(0,0)[br]{{\rotatebox{90}{Asymmetry $[10^{-4}]$}}}}
\put(-2,133){\makebox(0,0)[br]{{\rotatebox{90}{Asymmetry $[10^{-4}]$}}}}
\put(50,146){\mbox{\scriptsize${\cal R}_1^{tot}(b)$}}
\put(55,144.5){\mbox{$\vector(0,-1){5}$}}
\put(21,154.5){\mbox{\scriptsize$A_1^{FB}(b)$}}
\put(35.5,162.5){\mbox{\scriptsize$A_2^{FB}(b)$}}
\put(21,84){\mbox{\scriptsize$R_1^{FB}(b)$}}
\put(36,77){\mbox{\scriptsize$R_2^{FB}(b)$}}
\put(51,92){\mbox{\scriptsize$A^{FB}(b)$}}
\put(21,25.5){\mbox{\scriptsize$R_1^{FB}(l)$}}
\put(36,42){\mbox{\scriptsize$R_2^{FB}(l)$}}
\end{picture}
\end{center}
\caption{CP violating asymmetries (in units of $10^{-4}$)
  for the `mixed' scenario: $M=\mu=168$~GeV, $\tan\beta=4$,
  $m_{\tilde Q}=300$~GeV, $m_{\tilde U}=270$~GeV, $m_{\tilde D}=330$~GeV,
  $m_{\tilde\chi^+_1}\simeq 104$~GeV, and $m_{\tilde t_1}\simeq 96$~GeV.}
\label{fig:scen3}
\end{figure}


\begin{appendix}

\section{\boldmath Differential cross section for the $t$ decay
                   \label{app:polarization} }

\subsection{\boldmath The $t$ polarization four--vector}

Let us write the amplitude of the process \eq{1} in the form:
\begin{equation}
  M = \bar u(p_t)\,M_1\,u(-p_{\bar b})
\end{equation}
where
\begin{equation}
  M_1 = \frac{g^2}{2}\,\bar u(-p_{\bar d})\,
        \gamma_\alpha\,\frac{1-\gamma_5}{2}\,u(p_u)\,
        \frac{-i}{\hat s-m_W^2}\,\Gamma^\alpha \,.
\end{equation}

The polarization four--vector $\xi_\alpha$ of the $t$ quarks is
determined by
\begin{equation}
  \xi_{\alpha} =
  \left(g_{\alpha\beta}-\frac{p_\alpha^tp_\beta^t}{m_t^2}\right)\,
  \frac{Tr\left\{M_1\Lambda(-p_{\bar b})\bar M_1
                 \Lambda(p_t)\gamma^\beta\gamma^5\right\}}
       {Tr\left\{M_1\Lambda(-p_{\bar b})\bar M_1\Lambda(p_t)\right\}}
 \,, \quad \Lambda(p_t) = \not{\!p}_t + m_t \,.
\label{eq:xialpha}
\end{equation}
In the most general form the covariant decomposition of $\xi^\alpha$ reads
\begin{equation}
  \xi^\alpha =
  Q_1^\alpha {\cal P}_1^t + Q_2^\alpha {\cal P}_2^t
  + \varepsilon (\alpha,p_u,p_{\bar d},p_t){\cal D}^t.
\label{eq:xi}
\end{equation}
Here
\begin{equation}
  Q_1^\alpha = p_u^\alpha -\frac{(p_up_t)}{m_t^2}\,p_t^\alpha \,,\qquad
  Q_2^\alpha = p_{\bar d}^\alpha -\frac{(p_{\bar d}p_t)}{m_t^2}\,p_t^\alpha
\end{equation}
are two four--vectors in the production plane orthogonal to $p_t$, and
$\varepsilon (\alpha,p_u,p_{\bar d},p_t)$ is a four--vector orthogonal
to $p_t$ and to the production plane.
Using the notation
\begin{equation}
  {\cal N}(c_t) = a_0^{\bar b}+a_1^{\bar b}\,c_t + a_2^{\bar b}\,c_t^2 \,,
  \qquad c_t\equiv \cos\theta_t \,,
\end{equation}
and the usual Mandelstam variables
\begin{equation}
  \hat s = (p_u+p_{\bar d})^2\,,\quad
  \hat t = (p_u-p_{\bar b})^2 = (p_{\bar d}-p_t)^2\,,\qquad
  \hat s + \hat t + \hat u = m_t^2
\end{equation}
we obtain in the CMS of $u\bar d$:
\begin{eqnarray}
  {\cal P}_1^t
  &=& \frac{-1}{(m_t^2-\hat s)\,{\cal N}(c_t)}\,
      \frac{\Im m\,g_R^{CP}}{m_t} \left\{
      (m_t^2-\hat t)\,[4(\hat s-m_t^2) - 2(m_t^2-\hat u^2)-t\,]\right.\nn\\
  & & \hspace{60mm}\left.
    -\,\hat u(m_t^2-\hat u) + m_t^2(4\hat u+3\hat s) - \hat s^2\,\right\}\,,\\
  {\cal P}_2^t
  &=& \frac{1}{(m_t^2-\hat s )\,{\cal N}(c_t)}\,
      \left\{-4\,m_t\,\hat t\, (1-2\,\Im m\,f_L^{CP}) \right.\nn\\[3mm]
  & & \left. -\,\frac{\Im m\,g_R^{CP}}{m_t}
      \left[(m_t^2-\hat t)\,[-2(m_t^2-\hat u)-t] - \hat u(m_t^2 -\hat u)
            - m_t^2(4\hat t-3\hat s) - \hat s^2 \right] \right\} \,,\\
  {\cal D}^t
  &=& \frac{1}{(m_t^2-\hat s )\,{\cal N}(c_t)}\,
      \frac{4\,\Re e\,g_R^{CP}}{m_t}\,(\hat s -m_t^2)\,.
\end{eqnarray}

\subsection{General formula for the differential cross section}

From \eq{gen}, \eq{dgam} and \eq{xi} we obtain the general
formula for the differential cross section for the production of
the $t$ quarks and their subsequent decay in the CMS of $u\bar d$:
\begin{eqnarray}
  d\sigma_2^x
  &=& \frac{{\cal C}}{4\pi}\,
      \frac{m_t^2 Br(t\to xX}{E_t^2(1-\beta_t\cos\theta_{tx})^2}
      \left\{ {\cal N}^{SM} +  {\cal N}^{CP} \right.\nn\\
  & & +\,\alpha_x\,
      \frac{m_t\sqrt{\hat s}}{2E_t(1-\beta_t\cos\theta_{tx})}
      \left[(\tilde{\cal P}_+^{SM}+\tilde{\cal P}_+^{CP})
            \left(1-\frac{E_t^2}{m_t^2}(1-\beta_t\cos\theta_{tx})\right)
      \right.\nn\\
  & & -\,(\tilde{\cal P}_-^{SM} + \tilde{\cal P}_-^{CP})
      \left(\cos\theta_x-\beta_t\cos\theta_t
            \frac{E_t^2}{m_t^2}(1-\beta_t\cos\theta_{tx}) \right) \nn\\
  & & \left.\left.
      -\,\tilde{\cal D}\,\frac{\hat s}{2}\,\vert\vec p_t\vert\,
      (\vec n_u \vec n_t \vec n_x)\right]\right\}\,
      d\cos\theta_t\,d\cos\theta_x\,d\phi_x
\label{eq:general}
\end{eqnarray}
where we have used the notation
\begin{equation}
   \frac{\tilde{\cal P}_\pm}{{\cal N}(c_t)}=
{\cal P}_1^t \pm {\cal P}_2^t \,,
  \qquad
  \tilde{\cal P}_\pm = \tilde{\cal P}_\pm^{SM}+\tilde{\cal P}_\pm^{CP}\,,
  \qquad
  \frac{\tilde{\cal D}}{{\cal N}(c_t)}= {\cal D}^t \,,
\end{equation}
and
\begin{equation}
  {\cal N}(c_t)  = {\cal N}^{SM}(c_t) + {\cal N}^{CP}(c_t) \,,
\end{equation}
\begin{equation}
  E_t = \frac{\hat s +m_t^2}{2\sqrt{\hat s}} \,,\quad
  \vert\vec p_t\vert = \frac{\hat s -m_t^2}{2\sqrt{\hat s}} \,,\quad
  \beta_t = \frac{\hat s -  m_t^2}{\hat s + m_t^2} \,.
\end{equation}
$(\vec n_u \vec n_t \vec n_x)$ denotes the triple product
\begin{equation}
  (\vec n_u \vec n_t \vec n_x) =
  \vec{ n}_u \times \vec{ n}_t\cdot \vec{ n}_x
\end{equation}
with $\vec n_u$ the unit vector pointing in the direction of $\vec p_u$, etc.

\section{\boldmath The CP violating form factors in the MSSM
         \label{app:formfac}}


\subsection{Chargino--neutralino--stop loop}

The CP violating form factors from the
$\tilde t_n\,\tilde\chi^+_j\tilde\chi^0_k$ ($n,j = 1,2$; $k=1...4$)
loop are:
\begin{eqnarray}
   f_L^{CP}(\tilde\chi^+)
  &=& \frac{\alpha_w}{4\pi} \left\{ {\cal O}_1
      \left[ 2\,C_{00} - m_t^2\, C_{12} + \hat s \,(C_{22} + C_{2} + C_{12})
      \right] \right. \nn \\
  & & \left.  \hspace{8mm}
      +\,{\cal O}_2\, m_t \tilde m_k^0\, C_2
      - {\cal O}_3\, m_t \tilde m_j^+ (C_0 + C_2)
      - {\cal O}_4\, \tilde m_j^+ \tilde m_k^0 \,C_0 \right\}
  \label{fL}
\end{eqnarray}
and
\begin{equation}
   g_R^{CP}(\tilde\chi^+) = \frac{\alpha_w}{4\pi}\, m_t
  \left\{ {\cal O}_1 m_t \,C_{12} - {\cal O}_2 \tilde m_k^0 \,C_2
        + {\cal O}_3 \tilde m_j^+ (C_0 + C_1 + C_2) \right\} \,.
  \label{gR}
\end{equation}
Here we use the notation $\tilde m_j^+ = m_{\tilde\chi_j^+}$ and
$\tilde m_k^0 = m_{\tilde\chi_k^0}$.
The $C_X$ are the standard three--point functions \cite{pave} for which
we follow the convention of \cite{Denner}. In this case,
\begin{equation}
  C_X \equiv C_X(m_b^2,\, m_t^2,\, \hat s,\, m_{\tilde\chi_j^+}^2,\,
                     m_{\tilde t_n}^2,\, m_{\tilde\chi_k^0}^2)\,,
  \quad X \in \{0,1,2,00,11,12,22\}.
\end{equation}
See \app{CX} for the explicit definition of the $C_X$.
The couplings ${\cal O}_i \equiv {\cal O}_i^{\,njk} $ are
\begin{eqnarray}
  {\cal O}_1^{\,njk} &=& -\sqrt 2\,\,\Im m\,
                 (l_{nj}^{\tilde t}O^L_{kj}a^{\tilde t*}_{nk})\,,\\
  {\cal O}_2^{\,njk} &=& -\sqrt 2\,\,\Im m\,
                 (l_{nj}^{\tilde t}O^L_{kj}b^{\tilde t*}_{nk})\,,\\
  {\cal O}_3^{\,njk} &=& -\sqrt 2\,\,\Im m\,
                 (l_{nj}^{\tilde t}O^R_{kj}b^{\tilde t*}_{nk})\,,\\
  {\cal O}_4^{\,njk} &=& -\sqrt 2\,\,\Im m\,
                 (l_{nj}^{\tilde t}O^R_{kj}a^{\tilde t*}_{nk})\,,
\end{eqnarray}
with $l_{nj}^{\tilde t}$, $O^{L,R}_{kj}$, $a^{\tilde t}_{nk}$, and
$b^{\tilde t}_{nk}$ given in \app{mssm}.
Notice that in (\ref{fL}) and (\ref{gR}) one
has to sum over $n,j,k$.


\subsection{Gluino--stop--sbottom loop}

The $\tilde g\,\tilde t_n\tilde b_m$ loop gives:
\begin{equation}
   f_L^{CP}(\tilde g) = 0 \,,\label{fLgl}
\end{equation}
\begin{equation}
   g_R^{CP}(\tilde g) =
  -\frac{2}{3}\frac{\alpha_s}{\pi}\,m_t m_{\tilde g}\,
  {\cal O}_5\, C_1 \,,\label{gRgl}
\end{equation}
where
\begin{equation}
  C_1 \equiv C_1(m_b^2,\,m_t^2,\,\hat s,\,
            m_{\tilde b_m}^2,\,m_{\tilde g}^2,\,m_{\tilde t_n}^2)
\end{equation}
and
\begin{equation}
  {\cal O}_5 \equiv {\cal O}_5^{\,mn} = \vert \Gamma_{Lm}^{\tilde b}\vert^2\,
  \Im m\left(\Gamma_{Ln}^{\tilde t*}\Gamma_{Rn}^{\tilde t}\right)\,.
\end{equation}
With the explicit expression for the stop mixing matrix we have,
see \app{mssm},
\begin{eqnarray}
  \vert \Gamma_{Lm}^{\tilde b}\vert^2
    &=& \{\cos^2\theta_{\tilde b},\sin^2\theta_{\tilde b}\}\,,
        \qquad m=1,2,  \\
  \Im m \left(\Gamma_{Ln}^{\tilde t*}\Gamma_{Rn}^{\tilde t}\right)
    &=& \frac{1}{2}\sin\varphi_{\tilde t}\,\sin 2\theta_{\tilde t}\,
        \,\{1,-1\}\,,\qquad n=1,2\,.
\end{eqnarray}
Again, in (\ref{gRgl}) a summation over $m,n$ is assumed.


\subsection{Three point functions \label{app:CX}}

Here we give the definition of the Passarino--Veltman
three point functions~\cite{pave} used above in the
convention of \cite{Denner}.
For the general denominators we use the notation
\begin{equation}
  {\mathcal D}^{0} = q^{2} - m_{0}^{2}
  \quad \mbox{and}\quad
  {\mathcal D}^{j} = ( q + p_{j} )^{2} - m_{j}^{2}\,.
\end{equation}
Then the loop integrals in $D=4-\epsilon$ dimensions are as follows:
\begin{eqnarray}
  C_0 &=& \frac{1}{i\pi^2} \int d^{D}\! q \:
          \frac{1}{{\mathcal D}^0 {\mathcal D}^1 {\mathcal D}^2} \,,\\
  C_\mu &=& \frac{1}{i\pi^2} \int d^{D}\! q \:
      \frac{q_\mu}{{\mathcal D}^0 {\mathcal D}^1 {\mathcal D}^2}
      = p_{1\mu} C_1 + p_{2\mu} C_2  \,,\\
  C_{\mu\nu} &=& \frac{1}{i\pi^2} \int d^{D}\! q \:
      \frac{q_\mu q_\nu}{{\mathcal D}^0 {\mathcal D}^1 {\mathcal D}^2} \nn\\
      &=& g_{\mu\nu} C_{00} + p_{1\mu} p_{1\nu} C_{11}
          + ( p_{1\mu} p_{2\nu} + p_{2\mu} p_{1\nu} ) C_{12}
          + p_{2\mu} p_{2\nu} C_{22} \,.
\end{eqnarray}
where the $C$'s have
$(p_{1}^{2},(p_{1}-p_{2})^{2},p_{2}^{2},m_{0}^{2},m_{1}^{2},m_{2}^{2})$
as their arguments.

\section{Masses, Mixing Matrices, and Couplings \label{app:mssm}}

The neutralino mass matrix in the basis of
\be
\Psi_j^0=\left(-i\lambda ',-i\lambda^3,\psi_{H_1}^0,\psi_{H_2}^0\right)
\ee
is:
\begin{displaymath}
\cal{M^N}=
\left( \begin{array}{cccc}
M'&0 & -m_Z\sin\theta_W\cos\beta &m_Z\sin\theta_W\sin\beta \\
0& M & m_Z\cos\theta_W\cos\beta  &-m_Z\cos\theta_W\sin\beta  \\
-m_Z\sin\theta_W\cos\beta & m_Z\cos\theta_W\cos\beta  & 0 & -\mu \\
m_Z\sin\theta_W\sin\beta & - m_Z\cos\theta_W\sin\beta & -\mu &0
\end{array}\right)
\end{displaymath}
Here $\tan\beta = v_2/v_1$. This matrix is diagonalized by the
unitary neutralino mixing matrix $N$:
\be
 N^*{\cal{M^N}} N^\dagger
= {\cal M}^{\cal N}_D
\ee
where ${\cal M}^{\cal N}_D$ is a diagonal matrix with non-negative
elements --- $\tilde m_1^0, \tilde m_2^0, \tilde m_3^0, \tilde m_4^0$
--- the masses of the physical neutralino states.

The chargino mass matrix is:
\begin{displaymath}
\cal{M^C}=
\left( \begin{array}{cc}
M&\sqrt 2 m_W\sin\beta \\
\sqrt 2m_W\cos\beta & \mu
\end{array}\right)
\end{displaymath}
It is diagonalized by the two unitary matrices $U$ and $V$:
\be
U^*{\cal{M^C}}V^\dagger = {\cal M}^C_D,
\ee
where ${\cal M}^C_D$ is a diagonal matrix with non-negative entries
--- $\tilde m_1^+, \tilde m_2^+$ --- the masses of the physical
chargino states.

\noi
The mass matrix of the stops in the basis $(\tilde t_L,\,\tilde t_R)$ is
\begin{equation}
  {\cal M}_{\tilde t}^2 = \left(
  \begin{array}{cc}
    m_{\tilde Q}^2 + m_Z^2\cos 2\beta(\sfrac{1}{2}-\sfrac{2}{3}\sin^2\theta_W)
    + m_t^2 & (A_t^*-\mu\cot\beta)\,m_t \\
    (A_t-\mu\cot\beta)\,m_t
    & m_{\tilde U}^2 + \sfrac{2}{3}m_Z^2\cos 2\beta\sin^2\theta_W + m_t^2
  \end{array}\right) \,.
\end{equation}
${\cal M}_{\tilde t}^2$ is diagonalized by the rotation matrix
$\Gamma^{\tilde t}$ such that
$\Gamma^{\tilde t\,\dagger}\,{\cal M}_{\tilde t}^2\,\Gamma^{\tilde t} =
 diag(m_{\tilde t_1}^2,\,m_{\tilde t_2}^2)$ and
${\tilde{t}_{L} \choose \tilde{t}_{R}} =
 \Gamma^{\tilde t} {\tilde{t}_{1} \choose \tilde{t}_{2}}$.
We have:
\begin{equation}
\Gamma^{\tilde{t}} =
    \left(\begin{array}{rr}
      \Gamma^{\tilde{t}}_{L1}
    & \Gamma^{\tilde{t}}_{L2}
  \\  \Gamma^{\tilde{t}}_{R1}
    & \Gamma^{\tilde{t}}_{R2}
    \end{array}\right)
=   \left(\begin{array}{rr}
      e^{-\frac{i}{2} \varphi_{\tilde{t}}} \cos\theta_{\tilde{t}}
    & -e^{-\frac{i}{2} \varphi_{\tilde{t}}} \sin\theta_{\tilde{t}}
  \\ e^{\frac{i}{2} \varphi_{\tilde{t}}} \sin\theta_{\tilde{t}}
    & e^{\frac{i}{2} \varphi_{\tilde{t}}} \cos\theta_{\tilde{t}}
    \end{array}\right)
\; .
\end{equation}
The interaction Lagrangian which we need is:
\begin{eqnarray}
  {\cal L}_{b\tilde t_n\tilde \chi^+_j} \:\:
  &=& g\,\bar b\,(k_{nj}^{\tilde t}\,P_L + \,l_{nj}^{\tilde t}\,P_R)\,
      (\tilde {\chi}^+_j)^c \,\tilde t_n \,+\, {\rm h.c.} \,,\\
  {\cal L}_{t \tilde t_n  \tilde \chi^0_k} \:\:\,
  &=& g\,\bar t\,(b_{nk}^{\tilde t}\,P_L + a_{nk}^{\tilde t}\,P_R)
      \tilde\chi^0_k \,\tilde t_n \,+\, {\rm h.c.} \,,\\
  {\cal L}_{\tilde\chi^+_j\tilde \chi^0_k W}
  &=& g\,\bar{\tilde\chi}^0_k\, \gamma^\alpha \,
      (O_{kj}^L P_L + O_{kj}^R P_R)\, \tilde \chi^+_j \,W_{\alpha}^-
      \,+\, {\rm h.c.} \,.
\end{eqnarray}

\noi
The chargino--stop--bottom couplings are:
\begin{eqnarray}
  k_{nj}^{\tilde t} &=& +h_b\,U_{j2}^*\Gamma^{\tilde t}_{Ln} \,,\\
  l_{nj}^{\tilde t} &=& - V_{j1}\Gamma_{Ln}^{\tilde t}
                        + h_t V_{j2}\Gamma_{Rn}^{\tilde t}\,,
\end{eqnarray}
with
\begin{equation}
  h_t = \frac{m_t}{\sqrt 2\,m_W \sin\beta}\,, \qquad
  h_b = \frac{m_b}{\sqrt 2\,m_W\cos\beta}\,.
\end{equation}
The neutralino--stop--top couplings are:
\begin{eqnarray}
  a_{nk}^{\tilde t} &=& f_{Lk}^{\tilde t}\,\Gamma_{Ln}^{\tilde t}
                        + h_{Lk}^{\tilde t*}\,\Gamma_{Rn}^{\tilde t}\,,\\
  b_{nk}^{\tilde t} &=& h_{Lk}^{\tilde t}\,\Gamma^{\tilde t}_{Ln}
                        + f_{Rk}^{\tilde t}\,\Gamma_{Rn}^{\tilde t}\,,
\end{eqnarray}
with
\begin{eqnarray}
  f_{Lk}^{\tilde t} &=& -\sfrac{1}{\sqrt 2}\,(N_{k2}
                        +\sfrac{1}{3}\tan\theta_W\,N_{k1}) \,,\\
  f_{Rk}^{\tilde t} &=& \sfrac{2\sqrt 2}{3}\,\tan\theta_W\,N_{k1}^*\,,\\
  h_{Lk}^{\tilde t} &=& -h_t\, N_{k4}^*
\end{eqnarray}
in the basis $\Psi_j^0 = (-i\lambda',-i\lambda^3,\psi^0_{H_1},\psi^0_{H_2})$.
The chargino--neutralino--$W$ couplings are given by:
\begin{eqnarray}
  O_{kj}^L &=& -\sfrac{1}{\sqrt 2}\, N_{k4}\,V_{j2}^*
               + N_{k2}\,V_{j1}^* \,,\\
  O_{kj}^R &=& \sfrac{1}{\sqrt 2}\,N_{k3}^*\,U_{j2} + N_{k2}^*\,U_{j1}\,.
\end{eqnarray}

\end{appendix}
\newpage

\end{document}